\documentclass{amsart}

\usepackage[latin1]{inputenc}
\usepackage{amsmath, amsthm, amssymb, booktabs, multirow, graphicx, booktabs}
\usepackage{dialogue}
\usepackage[stable]{footmisc}
\usepackage{tikz,nicefrac}
\usetikzlibrary{calc,matrix,arrows,decorations.markings}
\usepackage[colorlinks]{hyperref}

\newtheorem{defin}{Definition}[section]

\newtheorem{theorem}[defin]{Theorem}

\newtheorem{definition}[defin]{Definition}

\newtheorem{example}[defin]{Example}

\begin{document}

\begin{abstract}
  This chapter is written for the forthcoming book ``A Concise
  Encyclopedia of Coding Theory'' (CRC press), edited by W.~Cary
  Huffman, Jon-Lark Kim, and Patrick Sol\'e. This book will collect
  short but foundational articles, emphasizing definitions, examples,
  exhaustive references, and basic facts on the model of the Handbook
  of Finite Fields. The target audience of the Encyclopedia is upper
  level undergraduates and graduate students.
\end{abstract}

\title{Semidefinite programming bounds for error-correcting codes}

\author{Frank Vallentin}
\address{F. Vallentin, Mathematisches Institut, Universit\"at zu K\"oln, Weyertal 86--90, 50931 K\"oln, Germany}
\email{frank.vallentin@uni-koeln.de}

\maketitle

\section{Introduction}
\label{sdpbounds:sec:intro}

Linear programming bounds belong to the most powerful and flexible
methods to obtain bounds for extremal problems in coding
theory. Initially, Delsarte \cite{Delsarte1973} developed linear
programming bounds in the algebraic framework of association schemes.

A central example in Delsarte's theory is finding upper bounds for the
parameter $A_2(n,d)$, the largest number of codewords in a binary code
of length $n$ with minimum Hamming distance $d$.

The application of linear programming bounds led to the best known
asymptotic bounds \cite{McElieceRRW1977}. It was realized that linear
programming bounds are also applicable to finite and infinite
two-point homogeneous spaces \cite[Chapter 9]{ConwayS1988}. These are
metric spaces in which the symmetry group acts transitively on pairs
of points having the same distance. So one can treat metric spaces
like the $q$-ary Hamming space $\mathbb{F}_q^n$, the sphere,
real/complex/quaternionic projective space, or Euclidean space
\cite{CohnE2003}.

In recent years, semidefinite programming bounds have been developed
with two aims: to strengthen linear programming bounds and to find
bounds for more general spaces. Semidefinite programs are convex
optimization problems which can be solved efficiently and which are a
vast generalization of linear programs. The optimization variable of a
semidefinite program is a positive semidefinite matrix whereas it is a
nonnegative vector for a linear program.

Schrijver \cite{Schrijver2005} was the first who applied semidefinite
programming bounds to improve the known upper bounds for $A_2(n, d)$
for many parameters $n$ and $d$. The underlying idea is that linear
programming bounds only exploit constraints involving pairs of
codewords, whereas semidefinite programming bounds can exploit
constraints between triples, quadruples, \dots\ of codewords.

This chapter introduces semidefinite programming bounds with an
emphasis on error-correcting codes. The structure of the chapter is as
follows:

In Section~\ref{sdpbounds:sec:sdp} the basic theory of linear
and semidefinite programming is reviewed in the framework of conic
programming.

Semidefinite programming bounds can be viewed as semidefinite
programming hierarchies for difficult combinatorial optimization
problems. One can express the computation of $A_2(n,d)$ as finding the
independence number of an appropriate graph $G(n,d)$ and apply the
Lasserre hierarchy to find upper bounds for $A_2(n,d)$. This approach is
explained in Section \ref{sdpbounds:sec:independent}.

The graph $G(n,d)$ has exponentially many vertices and the Lasserre
hierarchy for $G(n,d)$ employs matrices whose rows and columns are
indexed by all $t$-element subsets of $G(n,d)$, so a computation of
the semidefinite programs is not directly possible. However, the graph
has many symmetries and these symmetries can be exploited to
substantially reduce the size of the semidefinite programs. The
technique of symmetry reduction is the subject of Section
\ref{sdpbounds:sec:symmetry}. There, this technique is applied to the
graph $G(n,d)$ and the result of Schrijver is explained.

After Schrijver's breakthrough result, semidefinite programming bounds
were developed for different settings. These developments are reviewed
in Section~\ref{sdpbounds:sec:extensions}.

\section{Conic programming}
\label{sdpbounds:sec:sdp}

Semidefinite programming is a vast generalization of linear
programming. Geometrically, both linear and semidefinite programming
are concerned with minimizing or maximizing a linear functional over
the intersection of a fixed convex cone with an affine subspace. In
the case of linear programming the fixed convex cone is the
nonnegative orthant and the resulting intersection is a polyhedron. In
the case of semidefinite programming the fixed convex cone is the cone of
positive semidefinite matrices and the resulting intersection is a
spectrahedron. Linear and semidefinite programming belong to the field
of conic programming.

Textbooks and research monographs dealing with semidefinite
programming are: Wolkowicz, Saigal, and Vangenberghe (ed.)
\cite{WolkowiczSV2000}, Ben-Tal and Nemirovski \cite{Ben-TalN2001}, de
Klerk \cite{deKlerk2002}, Tun\c{c}el \cite{Tuncel2010},
Anjos and Lasserre (ed.) \cite{AnjosL2012}, G\"artner and Matou\v{s}ek
\cite{GaertnerM2012}, Blekherman, Parrilo and Thomas (ed.)
\cite{BlekhermanPT2013}, Laurent and Vallentin \cite{LaurentV2019}.

\subsection{Conic programming and its duality theory}
\label{sdpbounds:subsec:conicprogramming}

Conic programs are convex optimization problems. In general, conic
programming deals with minimizing or maximizing a linear functional
over the intersection of a fixed convex cone with an affine
subspace. See Nemirovski \cite{Nemirovski2007} and the references
therein for a detailed overview of conic programming.

Let $E$ be an $n$-dimensional real or complex vector space equipped
with a real-valued inner product $\langle \cdot, \cdot \rangle_E : E
\times E \to \mathbb{R}$.

\begin{definition}
  A set $K \subseteq E$ is called a \textbf{(convex) cone} if for all
  $x, y \in K$ and all nonnegative numbers $\alpha, \beta \in \mathbb{R}_+$ one has
  $\alpha x + \beta y \in K$. A convex cone $K$ is called
  \textbf{pointed} if $K \cap (-K) = \{0\}$. A convex cone is called
  \textbf{proper}, if it is pointed, closed, and full-dimensional. The
  \textbf{dual cone} of a convex cone $K$ is given by
\[
K^* = \{y \in E : \langle x, y \rangle_E \geq 0 \text{ for all } x \in
K\}.
\]
\end{definition}

The simplest convex cones are \textbf{finitely generated cones}; the
vectors $x_1, \ldots, x_N \in E$ determine the finitely generated cone
$K$ by
\[
K = \mathrm{cone}\{x_1, \ldots, x_N\} = \left\{\sum_{i=1}^N \alpha_i x_i :
\alpha_1, \ldots, \alpha_N \geq 0\right\}.
\]

A pointed convex cone $K \subseteq E$ determines a partial order on
$E$ by
\[
x \succeq_K y \text{ if and only if } x - y \in K.
\]

To define a conic program, we fix the space $E$, a proper convex cone
$K \subseteq E$, and an $m$-dimensional vector space $F$ with
inner product $\langle \cdot, \cdot \rangle_F$.

\begin{definition}
  A linear map $A \colon E \to F$ and vectors $c \in E$, $b \in F$
  determine a \textbf{primal conic program}
\[
p^* = \sup\{\langle c, x \rangle_E : x \in K, Ax = b\}.
\]
The corresponding \textbf{dual conic program} is
\[
d^* = \inf\{\langle b, y \rangle_F : y \in F, \overline{A}^{\sf T} y - c \in K^*\},
\]
where $\overline{A}^{\sf T} : F \to E$ is the usual adjoint of $A$.
\end{definition}

The vector $x \in E$ is the \textbf{optimization variable} of the
primal, the vector $y \in F$ is the optimization variable of the
dual. A vector $x$ is called \textbf{feasible} for the primal if
$x \in K$ and $Ax = b$. It is called \textbf{strictly feasible}
if additionally $x$ lies in the interior of $K$. It is called
\textbf{optimal} if $x$ is feasible and $p^* = \langle x,
c\rangle_E$.
Similarly, a vector $y$ is called feasible for the dual if
$\overline{A}^{\sf T} y - c \in K^*$, and it is called strictly feasible if
$\overline{A}^{\sf T} y - c$ lies in the interior of $K^*$. It is called optimal
if $y$ is feasible and $d^* = \langle b, y \rangle_F$.

The bipolar theorem (see for example Barvinok \cite{Barvinok2002} or
Simon \cite{Simon2011}) states that $(K^*)^* = K$ when $K$ is a proper
convex cone. From this it follows easily that taking the dual of the
dual conic program gives a conic program which is equivalent to the
primal.

Duality theory of conic programs looks at the (close) relationship
between the primal and dual conic programs. In particular duality can
be used to systematically find upper bounds for the primal program and
lower bounds for the dual program.

\begin{theorem}
(Duality theorem of conic programs)
\begin{enumerate}
\item \textbf{weak duality}: $p^* \leq d^*$.
\item \textbf{optimality condition/complementary slackness}: Suppose that $p^* = d^*$. Let $x$
  be a feasible solution for the primal and let $y$ be a feasible
  solution of the dual. Then $x$ is optimal for the primal and $y$ is
  optimal for the dual if and only if
  $\langle x, \overline{A}^{\sf T} y - c \rangle_E = 0$ holds.
\item \textbf{strong duality:} Suppose that primal and dual conic
  programs both have a strictly feasible solution. Then $p^* = d^*$
  and both primal and dual possess an optimal solution.
\end{enumerate}
\end{theorem}

\subsection{Linear programming}

To specialize conic programs to linear programs we choose $E$ to be
$\mathbb{R}^n$ with standard inner product
$\langle x,y \rangle_E = x^{\sf T} y$. For the convex cone $K$ we
choose the nonnegative orthant:

\begin{definition}
The \textbf{nonnegative orthant} is the following proper convex cone
\[
\mathbb{R}^n_+ = \{x \in \mathbb{R}^n : x_1 \geq 0, \ldots, x_n \geq 0\}.
\]
\end{definition}

The nonnegative orthant is self-dual,
$(\mathbb{R}^n_+)^* = \mathbb{R}^n_+$. So, for a matrix
$A \in \mathbb{R}^{m \times n}$, a vector $b \in \mathbb{R}^m$ and a
vector $c \in \mathbb{R}^n$ we get the \textbf{primal linear program}
\[
p^* = \sup\{c^{\sf T} x : x \geq 0, Ax = b\},
\]
and its \textbf{dual linear program}
\[
d^* = \inf\{b^{\sf T} y : y \in \mathbb{R}^m, A^{\sf T}y - c \geq 0\}.
\]
Here we simply write $x \geq 0$ for the partial order $x
\succeq_{\mathbb{R}^n_+} 0$.

Linear programming is a well established method, which is extremely
useful in theory and practice; see for example Schrijver
\cite{Schrijver1986}, Gr\"otschel, Lov\'asz, and Schrijver
\cite{GroetschelLS1988} and Wright \cite{Wright1997}. The main
algorithms to solve linear programs are the simplex method, the
ellipsoid method, and the interior-point method. Each one of these
three algorithms has specific advantages: In practice, the simplex
method and the interior-point method can solve very large
instances. The simplex method allows the computation of additional
information which is useful for the broader class of mixed integer
linear optimization problems, where some of the optimization variables
are constrained to be integers. The ellipsoid method and the
interior-point method are polynomial time algorithms. The ellipsoid
method is a versatile mathematical tool to prove the existence of
polynomial time algorithms, especially in combinatorial optimization.

\subsection{Semidefinite programming}

To specialize conic programs to semidefinite programs we choose $E$ to
be the $n(n+1)/2$-dimensional space $\mathcal{S}^n$ of real symmetric
$n \times n$ matrices. This space is equipped with the trace
(Frobenius) inner product
$\langle \cdot, \cdot \rangle_E = \langle \cdot, \cdot \rangle_T$
defined by
\[
\langle X, Y \rangle_T = \mathrm{Tr}(Y^{\sf T}X) = \sum_{i=1}^n \sum_{j=1}^n
X_{ij} Y_{ij}
\]
where $\mathrm{Tr}$ denotes the trace of a matrix. For the convex cone $K$ we
choose the cone of positive semidefinite matrices:

\begin{definition}
The \textbf{cone of positive semidefinite matrices} (or: the
\textbf{psd cone}) is the following proper convex cone:
\[
\mathcal{S}^n_+ = \{X \in \mathcal{S}^n : X \text{ is positive semidefinite}\}.
\]
\end{definition}

Let us recall that a matrix $X$ is positive semidefinite if and only
if for all $x \in \mathbb{R}^n$ we have $x^{\sf T}Xx \geq
0$. Alternatively, looking at a spectral decomposition of $X$, given by
\[
X = \sum_{i=1}^n \lambda_i u_i u_i^{\sf T},
\]
where $\lambda_i$ are the (real) eigenvalues of $X$ and $u_i$ is an
orthonormal basis consisting of corresponding eigenvectors, $X$ is
positive semidefinite if and only if all its eigenvalues are
nonnegative:
$\lambda = (\lambda_1, \ldots, \lambda_n) \in \mathbb{R}^n_+$. We
write $X \succeq 0$ for $X \succeq_{\mathcal{S}^n_+} 0$.

The cone of positive semidefinite matrices is self-dual,
$(\mathcal{S}^n_+)^* = \mathcal{S}^n_+$. So, for symmetric matrices
$A_1, \ldots, A_m \in \mathcal{S}^n$, a vector $b \in \mathbb{R}^m$ and a
symmetric matrix $C \in \mathcal{S}^n$ we get the \textbf{primal semidefinite program}
\begin{equation}
\label{sdpbounds:eq:primalsdp}
p^* = \sup\{ \langle C, X \rangle_T : X \succeq 0, \langle A_1, X
\rangle_T = b_1, \ldots, \langle A_m, X \rangle_T = b_m\}.
\end{equation}
Its dual semidefinite program is
\[
d^* = \inf\left\{ b^{\sf T} y : y \in \mathbb{R}^m, \sum_{j=1}^m y_j A_j - C \succeq 0\right\}.
\]
Restricting semidefinite programs to diagonal matrices, one recovers
linear programming as a special case of semidefinite programming.

\begin{definition}
The set of feasible solutions of a primal semidefinite program
\[
\mathcal{F} = \{X \in \mathcal{S}^n : X \succeq 0,  \langle A_j, X \rangle_T = b_j \text{ for } j = 1, \ldots, m\}
\]
is called a \textbf{spectrahedron}.
\end{definition}

Spectrahedra are generalizations of polyhedra. They are
central objects in convex algebraic geometry; see \cite{BlekhermanPT2013}.

Under mild technical assumptions one can solve semidefinite
programming problems in polynomial time. The following theorem was
proved in Gr\"otschel, Lov\'asz, and Schrijver \cite{GroetschelLS1988}
using the ellipsoid method and by de Klerk and Vallentin
\cite{deKlerkV2016} using the interior-point method.

\begin{theorem}
Consider the primal semidefinite program \eqref{sdpbounds:eq:primalsdp} with rational input $C$,
$A_1, \ldots, A_m$, and $b_1, \ldots, b_m$. Suppose we know a rational point
$X_0 \in \mathcal{F}$ and positive rational numbers $r$, $R$ so that
\[
B(X_0,r) \subseteq \mathcal{F} \subseteq B(X_0,R),
\]
where $B(X_0,r)$ is the ball of radius $r$, centered at $X_0$, in the
affine subspace
\[
\{X \in \mathcal{S}^n : \langle A_j, X \rangle_T = b_j \text{ for } j = 1,
\ldots, m\}.
\]
For every positive rational number $\epsilon > 0$ one can find in
polynomial time a rational matrix $X^* \in \mathcal{F}$ such that
\[
\langle C, X^* \rangle_T - p^* \leq \epsilon,
\]
where the polynomial is in $n$, $m$, $\log_2 \frac{R}{r}$,
$\log_2(1/\epsilon)$, and the bit size of the data $X_0$, $C$,
$A_1, \ldots, A_m$, and $b_1, \ldots, b_m$.
\end{theorem}

Sometimes---especially when dealing with invariant semidefinite
programs or in the area of quantum information theory---it is
convenient to work with complex Hermitian matrices instead of real
symmetric matrices. A complex matrix $X \in \mathbb{C}^{n \times n}$
is called Hermitian if $X = X^*$, where $X^* = \overline{X}^{\sf T}$
denotes the conjugate transpose of $X$,
i.e.~$X_{ij} = \overline{X}_{ji}$. A Hermitian matrix is called
positive semidefinite if for all vectors $x \in \mathbb{C}^n$ we have
$x^*X x \geq 0$. The space of Hermitian matrices is equipped with the
real-valued inner product $\langle X, Y \rangle_T =
\mathrm{Tr}(Y^*X)$. Now a \textbf{primal complex semidefinite program} is
\begin{equation}
\label{sdpbounds:eq:complexsdp}
p^* = \sup\{ \langle C, X \rangle_T : X \succeq 0, \langle A_1, X
\rangle_T = b_1, \ldots, \langle A_m, X \rangle_T = b_m\},
\end{equation}
where $A_1, \ldots, A_m \in \mathbb{C}^{n \times n}$, and $C \in
\mathbb{C}^{n \times n}$ are given Hermitian matrices, $b \in
\mathbb{R}^m$ is a given vector and $X \in \mathbb{C}^{n \times n}$ is
the positive semidefinite Hermitian optimization variable (denoted by
$X \succeq 0$).

One can easily reduce complex semidefinite programming to real
semidefinite programming by the following construction: A complex
matrix $X \in \mathbb{C}^{n \times n}$ defines a real matrix
\[
X' = 
\begin{pmatrix}
\Re(X) & -\Im(X)\\
\Im(X) & \Re(X)
\end{pmatrix} \in \mathbb{R}^{2n \times 2n},
\]
where $\Re(X) \in \mathbb{R}^{n \times n}$ and
$\Im(X) \in \mathbb{R}^{n \times n}$ are the real, respectively, the
imaginary parts of $X$. Then $X$ is Hermitian and positive
semidefinite if and only if $X'$ is symmetric and positive semidefinite.

\section{Independent sets in graphs}
\label{sdpbounds:sec:independent}

\subsection{Independence number and codes}

In the following we are dealing with finite simple graphs. These are
finite undirected graphs without loops and multiple edges. This means
that the vertex set is a finite set and the edge set consists of
(unordered) pairs of vertices.

\begin{definition}
  Let $G = (V, E)$ be simple finite graph with vertex set $V$ and edge
  set $E$. A subset of the vertices $I \subseteq V$ is called an
  \textbf{independent set} if every pair of vertices $x, y \in I$ is
  not adjacent, i.e.\ $\{x,y\} \not \in E$. The \textbf{independence
    number} $\alpha(G)$ is the largest cardinality of an independent
  set in $G$.
\end{definition}

In the optimization literature, independent sets are sometimes also
called \textbf{stable sets}, and the independence number is referred to as the
\textbf{stability number}.

\smallskip

Frequently the largest number of codewords in a code with given
parameters can be equivalently expressed as the independence number of
a specific graph.

\begin{example}
\label{sdpbounds:ex:graphgnd}
Recall that $A_2(n,d)$ is the largest number $M$ of codewords in a
binary code of length $n$ with minimum Hamming distance $d$. Consider
the graph $G(n,d)$ with vertex set $V = \mathbb{F}_2^n$ and edge set
$E = \{\{\mathbf{x}, \mathbf{y}\} : \mathrm{d}_{\mathrm{H}}(\mathbf{x}, \mathbf{y})
< d\}$.  Then independent sets in $G(n,d)$ are exactly binary codes
$\mathcal{C}$ of length $n$ with minimum Hamming distance
$d$. Furthermore, $A_2(n,d) = \alpha(G(n,d))$.

The graph $G(n,d)$ can also been seen as a Cayley graph over the
additive group $\mathbb{F}_2^n$. The vertices are the group elements
and two vertices $\mathbf{x}$ and $\mathbf{y}$ are adjacent if and
only if their difference $\mathbf{x} - \mathbf{y}$ has Hamming weight
strictly less than $d$.
\end{example}

Computing the independence number of a given a graph $G$ is generally
a very difficult problem. Computationally, determining even
approximate solutions of $\alpha(G)$ is an $\mathrm{NP}$-hard
problem; see H\aa{}stad \cite{Hastad1999}.

\subsection{Semidefinite programming bounds for the independence number}

One possibility to systematically find stronger and stronger upper
bounds for $\alpha(G)$, which is often quite good for graphs arising
in coding theory, is the Lasserre hierarchy of semidefinite
programming bounds.

The Lasserre hierarchy was introduced by Lasserre in
\cite{Lasserre2002}. He considered the general setting of $0/1$
polynomial optimization problems, and he proved that the hierarchy
converges in finitely many steps using Putinar's Positivstellensatz
\cite{Putinar1993}. Shortly after, Laurent \cite{Laurent2003} gave a
combinatorial proof, which we reproduce here.

The definition of the Lasserre hierarchy requires some notation. Let
$V$ be a finite set. By $\mathcal{P}_t(V)$ we denote the set of all
subsets of $V$ of cardinality at most $t$.

\begin{definition}
Let $t$ be an integer with $0 \leq t \leq n$. A symmetric matrix $M
\in \mathcal{S}^{\mathcal{P}_t(V)}$ is called a
\textbf{(combinatorial) moment matrix of order $t$} if
\[
M_{I,J} = M_{I',J'} \text{ whenever } I \cup J = I' \cup J'.
\]
A vector $y = (y_I) \in \mathbb{R}^{\mathcal{P}_{2t}(V)}$ defines a
combinatorial moment matrix of order $t$ by
\[
M_t(y) \in \mathcal{S}^{\mathcal{P}_t(V)} \quad \text{with} \quad
(M_t(y))_{I,J} = y_{I \cup J}.
\]
The matrix $M_t(y)$ is called the \textbf{(combinatorial) moment matrix of order
  $t$ of $y$}.
\end{definition}

\begin{example}
For $V = \{1,2\}$, the moment matrices of order one and order two of
$y$ have
the following form:
\[
M_1(y) =
\bordermatrix{
& \emptyset & 1 & 2 \cr
\emptyset & y_{\emptyset} & y_1 & y_2 \cr
1 & y_1 & y_1 & y_{12} \cr
2 & y_2 & y_{12} & y_{2}
}
\quad 
M_2(y) = 
\bordermatrix{ 
& \emptyset & 1 & 2 & 12 \cr
   \emptyset & y_{\emptyset} & y_1 & y_2 & y_{12} \cr
   1 & y_1 & y_1 & y_{12} & y_{12} \cr
   2 & y_2 & y_{12} & y_{2} & y_{12} \cr
   12& y_{12}& y_{12}& y_{12} & y_{12}
 }.
\]
Here and in the following, we simplify notation and use $y_i$ instead
of $y_{\{i\}}$ and $y_{12}$ instead of $y_{\{1,2\}}$.  Note that
$M_1(y)$ occurs as a principal submatrix of $M_2(y)$.
\end{example}

\begin{definition}
Let $G = (V, E)$ be a graph with $n$ vertices. Let $t$ be an integer
with $1 \leq t \leq n$. The \textbf{Lasserre bound of $G$ of order
 $t$} is the value of the semidefinite program
\begin{equation*}
\begin{split}
\mathrm{las}_t(G) = \max \Big\{\sum_{i \in V} y_i : \; & y \in
  \mathbb{R}^{\mathcal{P}_{2t}(V)}_+, \; y_{\emptyset}=1,\\[-3ex]
& \ y_{ij}=0 \text{ if } \{i,j\}\in E,\; M_t(y) \in \mathcal{S}^{\mathcal{P}_t(V)}_+\Big\}.
\end{split}
\end{equation*}
\end{definition}

\begin{theorem}
  The Lasserre bound of $G$ of order $t$ forms a hierarchy of stronger
  and stronger upper bounds for the independence number of $G$. In
  particular,
\[
\alpha(G) \leq \mathrm{las}_n(G) \leq \ldots \leq \mathrm{las}_2(G)\le
\mathrm{las}_1(G)
\]
holds.
\end{theorem}

\noindent\textbf{Proof:}
To show that $\alpha(G) \leq \mathrm{las}_t(G)$ for every
$1 \leq t \leq n$ we construct a feasible solution
$y \in \mathbb{R}^{\mathcal{P}_{2t}(V)}_+$ from any independent set
$I$ of $G$. This feasible solution will satisfy
$|I| = \sum_{i \in V} y_i$ and the desired inequality follows.  For
this, we simply set $y$ to be equal to the characteristic vector
$\chi^I \in \mathbb{R}^{\mathcal{P}_{2t}(V)}_+$ defined by
\[
\chi^I_J = 
\begin{cases}
1 & \text{if $J \subseteq I$,}\\
0 & \text{otherwise}.
\end{cases}
\]
Clearly, $y$ satisfies the conditions $y_\emptyset = 1$ and
$y_{ij} = 0$ if $i$ and $j$ are adjacent.  The moment matrix $M_t(y)$
is positive semidefinite because it is a rank-one matrix of the form
(note the slight abuse of notation here, $\chi^I$ is now interpreted
as a vector in $\mathbb{R}^{\mathcal{P}_{t}(V)}$)
\[
M_t(y) = \chi^I (\chi^I)^{\sf T} 
\;\text{where} \;\; M_t(y)_{J,J'} = y_{J \cup J'} = \chi^I_J
\chi^I_{J'} \;\; \text{and}
\;\;  \chi^I \in \mathbb{R}^{\mathcal{P}_{t}(V)}_+.
\]

Since $M_t(y)$ occurs as  a principal submatrix of $M_{t+1}(y)$, 
the inequality $\mathrm{las}_{t+1}(G)\le \mathrm{las}_t(G)$ follows.
\hfill$\Box$

One can show, using the \textbf{Schur complement} for block matrices,
\[
\text{for } A \text{ positive definite, then } 
\begin{pmatrix}
A & B\\
B^{\sf T} & C
\end{pmatrix} \succeq 0
\Longleftrightarrow C - B^{\sf T} A^{-1} B \succeq 0,
\]
that the first step of the Lasserre bound coincides with
the \textbf{Lov\'asz $\vartheta$-number} of $G$, a famous graph parameter which
Lov\'asz \cite{Lovasz1979} introduced to determine the Shannon capacity
$\Theta(C_5)$ of the cycle graph $C_5$. Determining the Shannon
capacity of a given graph is a very difficult problem and has
applications to the zero-error capacity of a noisy channel; see
Shannon \cite{Shannon1956}. For instance, the value of $\Theta(C_7)$
is currently not known.

\begin{theorem}
Let $G = (V, E)$ be a graph. We have $\mathrm{las}_1(G) =
\vartheta'(G)$ where $\vartheta'(G)$ is defined as the solution of the
following semidefinite program
\begin{equation*}
\begin{split}
\vartheta'(G) = \max\Big\{\sum_{i,j \in V} X_{i,j} \; : \; & X \in
\mathcal{S}^V_+,\; X_{i,j} \geq 0 \text{ for all } i,j \in V,\\[-3ex]
& \mathrm{Tr}(X) = 1,\; X_{i,j} = 0 \text{ if } \{i,j\} \in E \Big\}.
\end{split} 
\end{equation*}
\end{theorem}

Technically, the parameter $\vartheta'(G)$ is a slight variation of
the original Lov\'asz $\vartheta$-number as introduced in
\cite{Lovasz1979}. The difference is that in the definition of
$\vartheta(G)$ one omits the nonnegativity condition $X_{i,j} \geq 0$
for all $i,j \in V$.

Schrijver \cite{Schrijver1979} and independently McEliece, Rodemich,
and Rumsey \cite{McElieceRR1978} realized that $\vartheta'(G)$ is
nothing other than the Delsarte Linear Programming Bound in the special
case of the graph $G = G(n,d)$, which was defined in
Example~\ref{sdpbounds:ex:graphgnd}. We will provide a proof of this fact in
Section~\ref{sdpbounds:ssec:exdelsarte}.

\smallskip

An important feature of the Lasserre bound is that it does not loose
information. If the step of the hierarchy is high enough, we can
exactly determine the independence number of $G$.

\begin{theorem}
  For every graph $G$ the Lasserre bound of $G$ of order
  $t = \alpha(G)$ is exact; that means $\mathrm{las}_t(G) = \alpha(G)$
  for every $t \geq \alpha(G)$.
\end{theorem}

\noindent\textbf{Proof: (sketch)}
First we show that the hierarchy becomes stationary after $\alpha(G)$
steps.  Let $J \subseteq V$ be a set of vertices which contains an
edge, $\{i,j\} \in E$ with $i, j \in J$. Let
$y \in \mathbb{R}^{\mathcal{P}_{2t}(V)}$ be a feasible solution of
$\mathrm{las}_t(G)$ with $2t \geq |J|$. Then $y_J = 0$, which can be
seen as follows: Write $J = J_1 \cup J_2$ with $|J_1|, |J_2| \leq t$
and $\{i,j\} \subseteq J_1$. First, consider the following $2 \times 2$
principal submatrix of the positive semidefinite matrix $M_t(y)$
\[
\bordermatrix{
& ij & J_1 \cr
ij & y_{ij} & y_{J_1} \cr
J_1 & y_{J_1} & y_{J_1} \cr
}
\succeq 0 
\Longrightarrow y_{J_1} = 0,
\]
where we applied the constraint $y_{ij} = 0$. Then, consider the
following $2 \times 2$ principal submatrix of $M_t(y)$
\[
\bordermatrix{
& J_1 & J_2 \cr
J_1 & y_{J_1} & y_{J} \cr
J_2 & y_{J} & y_{J_2} \cr
}
\succeq 0 
\Longrightarrow y_{J} = 0.
\]
Hence, 
\[
\mathrm{las}_t(G) = \mathrm{las}_{t+1}(G) = \cdots = \mathrm{las}_n(G)
\quad \text{for } t \geq \alpha(G).
\]

The next step is showing that vectors
$y \in \mathbb{R}^{\mathcal{P}_n(V)}$, indexed by the full power set
$\mathcal{P}_n(V)$, which determine a positive semidefinite moment matrix
$M_n(y)$, form a finitely generated cone:
\[
M_n(y) \succeq 0 \Longleftrightarrow y \in \mathrm{cone}\{\chi^I : I
\subseteq V\},
\]
where $\chi^I$ are the characteristic vectors. Sufficiency follows
easily from $\chi^I_{J \cup J'} = \chi^I_J \chi^I_{J'}$. For
necessity, we first observe that the characteristic vectors form a
basis of $\mathbb{R}^{\mathcal{P}_n(V)}$. Let
$(\psi^J)_{J \in \mathcal{P}_n(V)}$ be its dual basis; it satisfies
$(\chi^I)^{\sf T} \psi^J = \delta_{I,J}$. Let $y$ be so that $M_n(y)$
is positive semidefinite and write $y$ in terms of the basis
\[
y = \sum_{I \in \mathcal{P}_n(V)} \alpha_I \chi^I \; \text{ with }
\alpha_I \in \mathbb{R}.
\]
Since $M_n(y)$ is positive semidefinite we have
\[
0 \leq (\psi^J)^{\sf T} M_n(y) \psi^J = \alpha_J.
\]

Now we finish the proof. Let $y \in \mathbb{R}^{\mathcal{P}_{n}(V)}$ be a
feasible solution of $\mathrm{las}_n(G)$. Then from the previous
arguments we see
\[
y = \sum_{I \text{ independent}} \alpha_I \chi^I, \text{ with }
\alpha_I \geq 0.
\]
Furthermore, the semidefinite program is normalized by
\[
1 = y_{\emptyset} = \sum_{I \text{ independent}} \alpha_I,
\]
and the objective value of $y$ equals
\[
\sum_{i \in V} y_i = \sum_{i \in V} \sum_{I \text{ indep.}}
\alpha_I \chi^I(i) = \sum_{I \text{ indep.}}
\alpha_I \sum_{i \in V} \chi^I(i) \leq 1 \cdot \alpha(G).
\]
\hfill$\Box$

\section{Symmetry reduction and matrix $*$-algebras}
\label{sdpbounds:sec:symmetry}

One can obtain semidefinite programming bounds for $A_2(n,d)$ by using
the Lasserre bound of order $t$ for the graph $G(n,d)$, defined in
Example~\ref{sdpbounds:ex:graphgnd}. Since the graph $G(n,d)$ has
exponentially many vertices, even computing the first step $t = 1$
amounts to solving a semidefinite program of exponential size. On the
other hand, the graph $G(n,d)$ is highly symmetric and these
symmetries can be used to simplify the semidefinite programs
considerably.

\subsection{Symmetry reduction of semidefinite programs}

Symmetry reduction of semidefinite programs is easiest explained using
complex semidefinite programs of the
form~\eqref{sdpbounds:eq:complexsdp}. Let $\Gamma$ be a finite group
and let $\pi : \Gamma \to \mathrm{U}(\mathbb{C}^n)$ be a
\textbf{unitary representation} of $\Gamma$; that is a group
homomorphism from $\Gamma$ to the group of unitary matrices
$\mathrm{U}(\mathbb{C}^n)$. Then $\Gamma$ acts on the space of complex
matrices by
\[
(g,X) \mapsto gX = \pi(g) X \pi(g)^*.
\]
A complex matrix $X$ is called \textbf{$\Gamma$-invariant} if $X = gX$
holds for all $g \in \Gamma$. By
\[
(\mathbb{C}^{n \times n})^{\Gamma} = \{X \in \mathbb{C}^{n \times n} : X = gX
\text{ for all } g \in \Gamma\}
\]
we denote the set of all $\Gamma$-invariant matrices.

\begin{definition}
  Let $\Gamma$ be a finite group. A complex semidefinite program is
  called \textbf{$\Gamma$-invariant} if for every feasible solution
  $X$ and every $g \in \Gamma$ the matrix $gX$ also is feasible and
  $\langle C,X\rangle_T = \langle C,gX \rangle_T$ holds. (Recall
  $\langle X, Y\rangle_T = \mathrm{Tr}(Y^*X)$.)
\end{definition}

Suppose that the complex semidefinite program~\eqref{sdpbounds:eq:complexsdp} is
$\Gamma$-invariant. Then we may restrict the optimization variable $X$ to
be $\Gamma$-invariant without changing the supremum. In fact, if $X$ is
feasible for~\eqref{sdpbounds:eq:complexsdp}, so is its \textbf{$\Gamma$-average}
\[
\overline{X} = \frac{1}{|\Gamma|} \sum_{g \in \Gamma} gX.
\]
Hence, \eqref{sdpbounds:eq:complexsdp} simplifies to
\begin{equation}
\label{sdpbounds:eq:complexsdp2}
\begin{split}
p^* = \sup\{ \langle C, X \rangle_T \; : \; &  X \succeq 0, X \in
(\mathbb{C}^{n \times n})^{\Gamma},\\
& \langle A_1, X \rangle_T = b_1, \ldots, \langle A_m, X \rangle_T = b_m\}.\\
\end{split}
\end{equation}
If we intersect the $\Gamma$-invariant complex matrices $(\mathbb{C}^{n
  \times n})^{\Gamma}$ with the Hermitian matrices we get a vector space
having a basis $B_1, \ldots, B_N$. If we express $X$ in terms of this
basis, \eqref{sdpbounds:eq:complexsdp2} becomes
\begin{equation}
\label{sdpbounds:eq:complexsdp3}
\begin{split}
p^* = \sup\{ \langle C, X \rangle_T \; : \; &  x_1, \ldots, x_N \in
\mathbb{C},\\
& X = x_1 B_1 + \cdots + x_N B_N \succeq 0,\\
& \langle A_1, X \rangle_T = b_1, \ldots, \langle A_m, X \rangle_T = b_m\}.\\
\end{split}
\end{equation}
So the number of optimization variables is $N$. It turns out that we
can simplify \eqref{sdpbounds:eq:complexsdp3} even more by performing a
simultaneous block diagonalization of the basis $B_1, \ldots,
B_N$.
This is a consequence of the main structure theorem of matrix
$*$-algebras.

\subsection{Matrix $*$-algebras}

\begin{definition}
A linear subspace $\mathcal{A} \subseteq \mathbb{C}^{n \times n}$ is
called a \textbf{matrix algebra} if it is closed under matrix
multiplication. It is called a \textbf{matrix $*$-algebra} if it is
closed under taking the conjugate transpose: if $A \in \mathcal{A}$,
then $A^* \in \mathcal{A}$.
\end{definition}

The space of $\Gamma$-invariant matrices $(\mathbb{C}^{n \times n})^{\Gamma}$ is a
  matrix $*$-algebra. Indeed, for $\Gamma$-invariant matrices $X, Y$ and
  $g \in \Gamma$, we have
\[
g(XY) = \pi(g)XY\pi(g)^* =(\pi(g)X\pi(g)^*)(\pi(g)Y\pi(g)^*) =
(gX)(gY) = XY,
\]
and
\[
g(X^*) = \pi(g)X^*\pi(g)^* = (\pi(g) X \pi(g)^*)^* = (gX)^* = X^*.
\]

The main structure theorem of matrix $*$-algebras---it is due to
Wedderburn and it is well-known in the theory of $C^*$-algebras, where
it can be also stated for the compact operators on a Hilbert
space---is the following:

\begin{theorem}
\label{sdpbounds:thm:matrixstarstructure}
Let $\mathcal{A} \subseteq \mathbb{C}^{n \times n}$ be a matrix
$*$-algebra. Then there are natural numbers $d$, $m_1, \ldots, m_d$
such that there is a $*$-isomorphism between $\mathcal{A}$ and a direct
sum of full matrix $*$-algebras
\[
\varphi \colon \mathcal{A} \to \bigoplus_{k = 1}^d \mathbb{C}^{m_k
  \times m_k}.
\]
Here a \textbf{$*$-isomorphism} is a bijective linear map between two
matrix $*$-algebras which respects multiplication and taking the
conjugate transpose.
\end{theorem}

An elementary proof of Theorem~\ref{sdpbounds:thm:matrixstarstructure}, which
also shows how to find a $*$-isomorphism $\varphi$ algorithmically, is
presented in~\cite{BachocGSV2012}. An alternative proof is given
in~\cite[Section 3]{Vallentin2009} in the framework of representation
theory of finite groups; see also \cite{Vallentin2008} and
\cite{Bachoc2009}.

Now we want to apply Theorem~\ref{sdpbounds:thm:matrixstarstructure} to block
diagonalize the $\Gamma$-invariant semidefinite
program~\eqref{sdpbounds:eq:complexsdp3}. Let
$\mathcal{A} = (\mathbb{C}^{n \times n})^{\Gamma}$ be the matrix $*$-algebra of
$\Gamma$-invariant matrices. Let $\varphi$ be a $*$-isomorphism as in
Theorem~\ref{sdpbounds:thm:matrixstarstructure}; then $\varphi$ preserves
positive semidefiniteness. Hence, \eqref{sdpbounds:eq:complexsdp3} is equivalent
to
\begin{equation*}
\begin{split}
p^* = \sup\{ \langle C, X \rangle_T \; : \; &  x_1, \ldots, x_N \in
\mathbb{C},\\
& x_1 \varphi(B_1) + \cdots + x_N \varphi(B_N) \succeq 0,\\
& X = x_1 B_1 + \cdots + x_N B_N,\\
& \langle A_1, X \rangle_T = b_1, \ldots, \langle A_m, X \rangle_T = b_m\}.\\
\end{split}
\end{equation*}
Thus, instead of dealing with one (potentially big) matrix of size $n
\times n$ one only has to work with $d$ (hopefully small) block
diagonal matrices of size $m_1, \ldots, m_d$. This reduces the
dimension from $n^2$ to $m_1^2 + \cdots + m_d^2$. Many practical
semidefinite programming solvers can take advantage of this block
structure and numerical calculations can become much faster. However,
finding an explicit $*$-isomorphism is usually a nontrivial task,
especially if one is interested in parameterized families of matrix
$*$-algebras.

\subsection{Example: The Delsarte Linear Programming Bound}
\label{sdpbounds:ssec:exdelsarte}

Let us apply the symmetry reduction technique to demonstrate that the
exponential size semidefinite program $\vartheta'(G(n,d))$ collapses
to the linear size Delsarte Linear Programming Bound.

Since the graph $G(n,d)$ is a Cayley graph over the additive group
$\mathbb{F}_2^n$, the semidefinite program $\vartheta'(G(n,d))$ is
$\mathbb{F}_2^n$-invariant where the group is acting as permutations
of the rows and columns of the matrix $X \in
\mathbb{C}^{\mathbb{F}_2^n \times \mathbb{F}_2^n}$. The graph $G(n,d)$
has even more symmetries. Its automorphism group
$\mathrm{Aut}(G(n,d))$ consists of all permutations of the $n$
coordinates $\mathbf{x} = x_1 x_2 \cdots x_n \in \mathbb{F}_2^n$
followed by independently switching the elements of $\mathbb{F}_2$
from $0$ to $1$, or vice versa. So the semidefinite program
$\vartheta'(G(n,d))$ is $\mathrm{Aut}(G(n,d))$-invariant. The
$*$-algebra $\mathcal{B}_n$ of $\mathrm{Aut}(G(n,d))$-invariant
matrices is called the \textbf{Bose-Mesner algebra (of the binary Hamming
  scheme)}. A basis $B_0, \ldots, B_n$ is given by zero-one matrices
\[
(B_r)_{\mathbf{x},\mathbf{y}} = 
\begin{cases}
1, & \text{if $\mathrm{d}_{\mathrm{H}}(\mathbf{x},\mathbf{y}) = r$,}\\
0, & \text{otherwise,}
\end{cases}
\]
with $r = 0, \ldots, n$. So, $\vartheta'(G(n,d))$ in the form
of~\eqref{sdpbounds:eq:complexsdp3} is the following semidefinite program in
$n+1$ variables:
\[
\begin{split}
\max\Big\{
2^n \sum_{r=0}^n \binom{n}{r} x_r  : \; & x_0 =
\frac{1}{2^n}, \; x_1 = \cdots = x_{d-1} = 0,\\
&  x_d, \ldots, x_n \geq 0, \; \sum_{r=0}^n x_r B_r \succeq 0 \Big\}.
\end{split}
\]
Finding a simultaneous block diagonalization of the $B_r$'s is easy
since they pairwise commute and have a common system of
eigenvectors. An orthogonal basis of eigenvectors is given by
$\chi_{\mathbf{a}} \in \mathbb{C}^{\mathbb{F}_2^n}$ defined
componentwise by
\[
(\chi_{\mathbf{a}})_{\mathbf{x}} = \prod_{j=1}^n (-1)^{a_j x_j}.
\]
Indeed,
\[
\begin{split}
(B_r \chi_\mathbf{a})_{\mathbf{x}} \; = \; & 
\sum_{\mathbf{y} \in \mathbb{F}_2^n} (B_r)_{\mathbf{x},\mathbf{y}}
(\chi_{\mathbf{a}})_{\mathbf{y}} \\
\; = \; & 
\sum_{\mathbf{y} \in \mathbb{F}_2^n} (B_r)_{\mathbf{x},\mathbf{y}}
(\chi_{\mathbf{a}})_{\mathbf{y - x}} (\chi_{\mathbf{a}})_{\mathbf{x}}\\
 = &
\left(\sum_{\mathbf{y} \in \mathbb{F}_2^n, \mathrm{d}_{\mathrm{H}}(\mathbf{x},\mathbf{y}) = r}
  (\chi_{\mathbf{a}})_{\mathbf{y} - \mathbf{x}} \right)
(\chi_\mathbf{a})_{\mathbf{x}}\\
= &
\left(\sum_{\mathbf{y} \in \mathbb{F}_2^n, \mathrm{d}_{\mathrm{H}}(\mathbf{0},\mathbf{y}) = r}
  (\chi_{\mathbf{a}})_{\mathbf{y}} \right)
(\chi_\mathbf{a})_{\mathbf{x}}.
\end{split}
\]
The eigenvalues are given by the \textbf{Krawtchouk polynomials}
\[
K^{(n,2)}_r(x) = \sum_{j=0}^r (-1)^j \binom{x}{j} \binom{n-x}{r-j}
\]
through
\[
\sum_{\mathbf{y} \in \mathbb{F}_2^n, \mathrm{d}_{\mathrm{H}}(\mathbf{0},\mathbf{y})
  = r}
  (\chi_{\mathbf{a}})_{\mathbf{y}} = K^{(n,2)}_r(\mathrm{d}_{\mathrm{H}}(\mathbf{0},\mathbf{a})).
\]
Altogether, we have the $*$-algebra isomorphism
\[
\varphi : \mathcal{B}_n \to \bigoplus_{r=0}^n \mathbb{C},
\]
(so $m_0 = \cdots = m_n = 1$) defined by
\[
\varphi(B_r) = (K^{(n,2)}_r(0), K^{(n,2)}_r(1), \ldots, K^{(n,2)}_r(n)).
\]
So the semidefinite program $\vartheta'(G(n,d))$ degenerates to
the following linear program
\[
\begin{split}
\max\Big\{
2^n \sum_{r=0}^n \binom{n}{r} x_r  : \; & \; x_0 =
\frac{1}{2^n}, \; x_1 = \cdots = x_{d-1} = 0, \; x_d, \ldots, x_n \geq
0,\\
& \; \sum_{r=0}^n x_r K^{(n,2)}_r(j) \geq 0 \text{ for } j  = 0, \ldots, n\Big\}.
\end{split}
\]
This is the Delsarte Linear Programming Bound.

\subsection{Example: The Schrijver Semidefinite Programming Bound}
\label{sdpbounds:sec:sdpbounds}

To set up a stronger semidefinite programming bound one can apply the
Lasserre bound directly, but also many variations are possible. These
variations are crucial to be able to exploit the symmetries of the
problem at hand. For instance, one can consider only ``interesting''
principal submatrices of the moment matrices to simplify the
computation.

A rough classification for these variations can be given in terms of
\textit{$k$-point bounds}. This refers to all variations which make
use of variables $y_I$ with $|I| \leq k$. A $k$-point bound is capable
of using obstructions coming from the local interaction of
configurations having at most $k$ points. For instance Lov\'asz
$\vartheta$-number is a $2$-point bound and the $t$-th step in
Lasserre's hierarchy is a $2t$-point bound. The relation between
$k$-point bounds and Lasserre's hierarchy was first made explicit by
Laurent~\cite{Laurent2007} in the case of bounds for binary codes; see
also Gijswijt \cite{Gijswijt2009}, who discusses the symmetry
reduction needed to compute $k$-point bounds for block codes, and de
Laat and Vallentin \cite{deLaatV2015}, who consider $k$-point bounds
for compact topological packing graphs.

Schrijver's bound for binary codes~\cite{Schrijver2005} is a $3$-point
bound. Essentially, it looks at principal submatrices $M_{\mathbf{a}}
\in \mathbb{R}^{\mathbb{F}_2^n \times \mathbb{F}_2^n}$ of the matrix
$M_2(y)$ defined by
\[
(M_{\mathbf{a}}(y))_{\mathbf{b},\mathbf{c}} =
y_{\{\mathbf{a},\mathbf{b},\mathbf{c}\}} \quad \text{with }
  \mathbf{a},\mathbf{b},\mathbf{c} \in \mathbb{F}_2^n.
\]
The group which leaves the corresponding semidefinite program
invariant is the stabilizer of a codeword in $\mathrm{Aut}(G(n,d))$,
which is the symmetric group permuting the $n$ coordinates of
$\mathbb{F}_2^n$.

The algebra $\mathcal{A}_n \subseteq \mathbb{R}^{\mathbb{F}_2^n \times
  \mathbb{F}_2^n}$ invariant under this group action is called the
\textbf{Terwilliger algebra of the binary Hamming scheme}. Schrijver
determined a block diagonalization of the Terwilliger algebra which we
recall here.

For nonnegative integers $i,j,t$, with $t \leq i, j$ and $i + j \leq n
+ t$, the matrices
\[
(B^t_{i,j})_{\mathbf{x}, \mathbf{y}} =
\begin{cases}
  1, & \text{ if $\mathrm{wt}_{\mathrm{H}}(\textbf{x}) = i$, $\mathrm{wt}_{\mathrm{H}}(\textbf{y}) = j$, $\mathrm{d}_{\mathrm{H}}(\mathbf{x},\mathbf{y}) = i + j - 2t$,}\\
  0, & \text{ otherwise.}
\end{cases}
\]
form a basis of~$\mathcal{A}_n$. Hence, $\dim \mathcal{A}_n = \binom{n+3}{3}$. 
The desired $*$-isomorphism
\[
\varphi : \mathcal{A}_n \to \bigoplus_{k=0}^{\lfloor n/2 \rfloor} \mathbb{C}^{(n-2k+1) \times (n-2k+1)}
\]
is defined as follows: Set
\[
\beta^t_{i,j,k}
=
\sum_{u=0}^n (-1)^{u-t} \binom{u}{t} \binom{n-2k}{u-k} \binom{n-k-u}{i-u} \binom{n-k-u}{j-u}
\]
so that
\[
\varphi(B^t_{i,j}) =
\left(\ldots,
\begin{pmatrix}
  \binom{n-2k}{i-k}^{-1/2} \binom{n-2k}{j-k}^{-1/2} \beta^t_{i,j,k}
\end{pmatrix}^{n-k}_{i,j = k},
\ldots\right)_{k = 0, \ldots, \lfloor n/2 \rfloor}.
\]
Schrijver determined the $*$-isomorphism from first principles using
linear algebra. Later, Vallentin~\cite{Vallentin2009} used
representation theory of finite groups to derive an alternative
proof. Here the connection to the orthogonal Hahn and Krawtchouk
polynomials becomes visible. Another constructive proof of the
explicit block diagonalization of $\mathcal{A}_n$ was given by
Srinivasan~\cite{Srinivasan2011}; see also Martin and Tanaka
\cite{MartinT2009}.

\section{Extensions and ramifications}
\label{sdpbounds:sec:extensions}

Explicit computations of $k$-point semidefinite programming bounds
have been done in a variety of situations, in the finite and infinite
setting. Table~\ref{sdpbounds:table:kpointbounds} gives a guide to the
literature.

\begin{table}[htb]
\begin{center}
\caption{Computation of $k$-point bounds.}
\label{sdpbounds:table:kpointbounds}
\medskip
\scriptsize
\begin{tabular}{@{}llll@{}}
\bf Problem & \bf $2$-point bound & \bf $3$-point bound & \bf $4$-point bound\\[1ex]

Binary codes & Delsarte \cite{Delsarte1973} & Schrijver \cite{Schrijver2005} & \parbox{0.15\textwidth}{Gijswijt,\\ Mittelmann,\\ Schrijver \cite{GijswijtMS2012}}\\[4ex]

$q$-ary codes & Delsarte \cite{Delsarte1973}
                                          & \parbox{0.17\textwidth}{Gijswijt,\\
  Schrijver,\\ Tanaka \cite{GijswijtST2006}} & 
\parbox{0.17\textwidth}{Litjens,\\ Polak,\\ Schrijver \cite{LitjensPS2017}}
\\[4ex]

Constant weight codes & Delsarte \cite{Delsarte1973} & \parbox{0.17\textwidth}{Schrijver \cite{Schrijver2005},\\  Regts \cite{Regts2009}} & Polak \cite{Polak2019}\\[3ex]

Lee codes & Astola \cite{Astola1982} & Polak \cite{Polak2018} \\[2ex]

Bounded weight codes & \parbox{0.17\textwidth}{Bachoc,\\Chandar,\\Cohen,\\Sol\'e\\Tchamkerten \cite{BachocCCST2011}}\\[6ex]

Grassmannian codes & Bachoc \cite{Bachoc2006}\\[2ex]

Projective codes & \parbox{0.17\textwidth}{Bachoc,\\Passuello,\\Vallentin \cite{BachocPV2013}} \\[4ex]

Spherical codes & \parbox{0.17\textwidth}{Delsarte,\\Goethals,\\Seidel \cite{DelsarteGS1977}} & \parbox{0.16\textwidth}{Bachoc,\\Vallentin \cite{BachocV2008}} &  \\[4ex]

Codes in $\mathbb{R}\mathrm{P}^{n-1}$
                    & \parbox{0.17\textwidth}{Kabatiansky,\\Levenshtein
  \cite{KabatianskyL1978}}& \parbox{0.16\textwidth}{Cohn,\\Woo
  \cite{CohnW2012}} & \\[3ex]

Sphere packings & \parbox{0.17\textwidth}{Cohn,\\ Elkies \cite{CohnE2003}}&  \\[3ex]

\parbox{0.22\textwidth}{Binary sphere and\\ spherical cap packings}
& \parbox{0.17\textwidth}{de Laat, \\Oliveira,\\ Vallentin
  \cite{deLaatOV2014}}& \\[4ex]

Translative body packings & \parbox{0.17\textwidth}{Dostert,\\
  Guzm\'an,\\Oliveira,\\Vallentin \cite{DostertGOV2017}}&  \\[5ex]

\parbox{0.22\textwidth}{Congruent copies\\ of a convex body}
& \parbox{0.17\textwidth}{Oliveira,\\ Vallentin
  \cite{OliveiraV2018}}& \\
\end{tabular}
\smallskip
\end{center}
\end{table}

Semidefinite programming bounds have also been developed for
generalized distances and list decoding radii of binary codes by
Bachoc and Z\'emor \cite{BachocZ2010}, for permutation codes by
Bogaerts and Dukes \cite{BogaertsD2014}, for mixed binary/ternary
codes by Litjens \cite{Litjens2018}, for subsets of coherent
configurations by Hobart \cite{Hobart2009} and Hobart and Williford
\cite{HobartW2014}, for ordered codes by Trinker \cite{Trinker2011},
for energy minimization on $S^2$ by de Laat \cite{deLaat2016} and for
spherical two-distance sets and for equiangular lines by Barg and Yu
\cite{BargY2013} and by Machado, de Laat, Oliveira, and
Vallentin~\cite{deLaatMOV2018}. They have been used by Brouwer and
Polak \cite{BrouwerP2017} to prove the uniqueness of several constant
weight codes.

In extremal combinatorics, (weighted) vector space versions of the
Erd\H{o}s-Ko-Rado Theorem for cross intersecting families have been
proved using semidefinite programming bounds by Suda and Tanaka
\cite{SudaT2014} and by Suda, Tanaka and Tokushige \cite{SudaTT2017},
see also the survey by Frankl and Tokushige \cite{FranklT2016}.

Another coding theory application of the symmetry reduction technique
are new approaches to the Assmus-Mattson Theorem by Tanaka
\cite{Tanaka2009} and by Morales and Tanaka \cite{MoralesT2018}

\section*{Acknowledgements}

The author was partially supported by the SFB/TRR 191 ``Symplectic
Structures in Geometry, Algebra and Dynamics'', funded by the DFG.  He
also gratefully acknowledges support by DFG grant VA 359/1-1. This
project has received funding from the European Unions Horizon 2020
research and innovation programme under the Marie Sk\l{}odowska-Curie
agreement number 764759.


\begin{thebibliography}{10}

\bibitem{AnjosL2012}
M.~F. Anjos and J.~B. Lasserre, editors.
\newblock {\em {Handbook on Semidefinite, Conic and Polynomial Optimization}}.
\newblock Springer-Verlag, New York, 2012.

\bibitem{Astola1982}
J.~Astola.
\newblock The {L}ee-scheme and bounds for {L}ee-code.
\newblock {\em Cybernetics and Systems}, 13:331--343, 1982.

\bibitem{Bachoc2006}
C.~Bachoc.
\newblock Linear programming bounds for codes in {G}rassmannian spaces.
\newblock {\em IEEE Trans. Inform. Theory}, IT--52:2111--2125, 2006.

\bibitem{Bachoc2009}
C.~Bachoc.
\newblock Semidefinite programming, harmonic analysis and coding theory.
\newblock arXiv:0909.4767 [cs.IT], 2009.

\bibitem{BachocCCST2011}
C.~Bachoc, V.~Chandar, G.~Cohen, P.~Sol\'{e}, and A.~Tchamkerten.
\newblock On bounded weight codes.
\newblock {\em IEEE Trans. Inform. Theory}, IT--57:6780--6787, 2011.

\bibitem{BachocGSV2012}
C.~Bachoc, D.C. Gijswijt, A.~Schrijver, and F.~Vallentin.
\newblock Invariant semidefinite programs.
\newblock In M.~F. Anjos and J.~B. Lasserre, editors, {\em Handbook on
  Semidefinite, Conic and Polynomial Optimization}, pages 219--269.
  Springer-Verlag, New York, 2012.

\bibitem{BachocPV2013}
C.~Bachoc, A.~Passuello, and F.~Vallentin.
\newblock {Bounds for projective codes from semidefinite programming}.
\newblock {\em Adv. Math. Comm.}, 7:127--145, 2013.

\bibitem{BachocV2008}
C.~Bachoc and F.~Vallentin.
\newblock New upper bounds for kissing numbers from semidefinite programming.
\newblock {\em J. Amer. Math. Soc.}, 21:909--924, 2008.

\bibitem{BachocZ2010}
C.~Bachoc and G.~Z\'{e}mor.
\newblock Bounds for binary codes relative to pseudo-distances of {$k$} points.
\newblock {\em Adv. Math. Commun.}, 4:547--565, 2010.

\bibitem{BargY2013}
A.~Barg and W.-H. Yu.
\newblock New bounds for spherical two-distance sets.
\newblock {\em Exp. Math.}, 22:187--194, 2013.

\bibitem{Barvinok2002}
A.~Barvinok.
\newblock {\em {A course in convexity}}.
\newblock American Mathematical Society, Providence, RI, 2002.

\bibitem{Ben-TalN2001}
A.~Ben-Tal and A.~Nemirovski.
\newblock {\em Lectures on modern convex optimization}.
\newblock Society for Industrial and Applied Mathematics (SIAM), Philadelphia,
  PA; Mathematical Programming Society (MPS), Philadelphia, PA, 2001.

\bibitem{BlekhermanPT2013}
G.~Blekherman, P.~A. Parrilo, and R.~R. Thomas, editors.
\newblock {\em Semidefinite optimization and convex algebraic geometry}.
\newblock Society for Industrial and Applied Mathematics (SIAM), Philadelphia,
  PA; Mathematical Optimization Society, Philadelphia, PA, 2013.

\bibitem{BogaertsD2014}
M.~Bogaerts and P.~Dukes.
\newblock Semidefinite programming for permutation codes.
\newblock {\em Discrete Math.}, 326:34--43, 2014.

\bibitem{BrouwerP2017}
A.~E. Brouwer and S.~C. Polak.
\newblock Uniqueness of codes using semidefinite programming.
\newblock {\em Des. Codes Cryptogr.}, 2018.

\bibitem{CohnE2003}
H.~Cohn and N.~Elkies.
\newblock New upper bounds on sphere packings {I}.
\newblock {\em Ann. of Math.}, 157:689--714, 2003.

\bibitem{CohnW2012}
H.~Cohn and J.~Woo.
\newblock Three-point bounds for energy minimization.
\newblock {\em J. Amer. Math. Soc.}, 25:929--958, 2012.

\bibitem{ConwayS1988}
J.~H. Conway and N.~J.~A Sloane.
\newblock {\em Sphere Packings, Lattices, and Groups}.
\newblock Springer-Verlag, New York, 1988.

\bibitem{deKlerk2002}
E.~de~Klerk.
\newblock {\em Aspects of semidefinite programming}.
\newblock Kluwer Academic Publishers, Dordrecht, 2002.

\bibitem{deKlerkV2016}
E.~de~Klerk and F.~Vallentin.
\newblock On the {T}uring model complexity of interior point methods for
  semidefinite programming.
\newblock {\em SIAM J. Optim.}, 26:1944--1961, 2016.

\bibitem{deLaat2016}
D.~de~Laat.
\newblock Moment methods in energy minimization: New bounds for riesz minimal
  energy problems.
\newblock arXiv:1610.04905 [math.OC], 2016.

\bibitem{deLaatOV2014}
D.~de~{L}aat, F.M. de~Oliveira~Filho, and F.~Vallentin.
\newblock Upper bounds for packings of spheres of several radii.
\newblock {\em Forum of Mathematics, Sigma}, 2:42 pages, 2014.

\bibitem{deLaatMOV2018}
D.~de~Laat, F.~C. Machado, F.~M. de~Oliveira~Filho, and F.~Vallentin.
\newblock $k$-point semidefinite programming bounds for equiangular lines.
\newblock arXiv:1812.06045 [math.OC], 2018.

\bibitem{deLaatV2015}
D.~de~Laat and F.~Vallentin.
\newblock A semidefinite programming hierarchy for packing problems in discrete
  geometry.
\newblock {\em Math. Program., Ser. B}, 151:529--553, 2015.

\bibitem{OliveiraV2018}
F.M. de~Oliveira~Filho and F.~Vallentin.
\newblock Computing upper bounds for packing densities of congruent copies of a
  convex body.
\newblock In G.~Ambrus, I.~B\'ar\'any, K.J. B\"or\"oczky, G.~Fejes~T\'oth, and
  J.~Pach, editors, {\em New Trends in Intuitive Geometry}, pages 155--188.
  Springer, 2018.

\bibitem{Delsarte1973}
P.~Delsarte.
\newblock An algebraic approach to the association schemes of coding theory.
\newblock {\em Philips Res. Rep. Suppl.}, 1973.

\bibitem{DelsarteGS1977}
P.~Delsarte, J.-M. Goethals, and J.~J. Seidel.
\newblock Spherical codes and designs.
\newblock {\em Geometriae Dedicata}, 6:363--388, 1977.

\bibitem{DostertGOV2017}
M.~Dostert, C.~Guzman, F.~M. de~Oliveira~Filho, and F.~Vallentin.
\newblock New upper bounds for the density of translative packings of
  three-dimensional convex bodies with tetrahedral symmetry.
\newblock {\em Discrete {\&} Computational Geometry}, 58:449--481, 2017.

\bibitem{FranklT2016}
P.~Frankl and N.~Tokushige.
\newblock Invitation to intersection problems for finite sets.
\newblock {\em J. Combin. Theory Ser. A}, 144:157--211, 2016.

\bibitem{GaertnerM2012}
B.~G\"{a}rtner and J.~Matou\v{s}ek.
\newblock {\em Approximation algorithms and semidefinite programming}.
\newblock Springer, Heidelberg, 2012.

\bibitem{Gijswijt2009}
D.~Gijswijt.
\newblock Block diagonalization for algebras associated with block codes.
\newblock arXiv:0910.4515 [math.OC], 2009.

\bibitem{GijswijtST2006}
D.~Gijswijt, A.~Schrijver, and H.~Tanaka.
\newblock {New upper bounds for nonbinary codes based on the Terwilliger
  algebra and semidefinite programming}.
\newblock {\em Journal of Combinatorial Theory, Series A}, 113:1719--1731,
  2006.

\bibitem{GijswijtMS2012}
D.~C. Gijswijt, H.~D. Mittelmann, and A.~Schrijver.
\newblock Semidefinite code bounds based on quadruple distances.
\newblock {\em IEEE Trans. Inform. Theory}, IT--58:2697--2705, 2012.

\bibitem{GroetschelLS1988}
M.~Gr\"{o}tschel, L.~Lov\'{a}sz, and A.~Schrijver.
\newblock {\em Geometric algorithms and combinatorial optimization}.
\newblock Springer-Verlag, Berlin, 1988.

\bibitem{Hastad1999}
J.~H\aa{}stad.
\newblock Clique is hard to approximate within {$n^{1-\epsilon}$}.
\newblock {\em Acta Math.}, 182:105--142, 1999.

\bibitem{Hobart2009}
S.~A. Hobart.
\newblock Bounds on subsets of coherent configurations.
\newblock {\em Michigan Math. J.}, 58:231--239, 2009.

\bibitem{HobartW2014}
S.~A. Hobart and J.~Williford.
\newblock The absolute bound for coherent configurations.
\newblock {\em Linear Algebra Appl.}, 440:50--60, 2014.

\bibitem{KabatianskyL1978}
G.~A. Kabatiansky and V.~I. Levenshtein.
\newblock On bounds for packings on a sphere and in space.
\newblock {\em Probl. Peredachi Inf.}, 14:3--25, 1978.

\bibitem{Lasserre2002}
J.~B. Lasserre.
\newblock An explicit equivalent positive semidefinite program for nonlinear
  {$0$}-{$1$} programs.
\newblock {\em SIAM J. Optim.}, 12:756--769, 2002.

\bibitem{Laurent2003}
M.~Laurent.
\newblock A comparison of the {S}herali-{A}dams, {L}ov\'{a}sz-{S}chrijver, and
  {L}asserre relaxations for 0-1 programming.
\newblock {\em Math. Oper. Res.}, 28:470--496, 2003.

\bibitem{Laurent2007}
M.~Laurent.
\newblock Strengthened semidefinite programming bounds for codes.
\newblock {\em Math. Program.}, 109:239--261, 2007.

\bibitem{LaurentV2019}
M.~Laurent and F.~Vallentin.
\newblock {\em A course on semidefinite optimization}.
\newblock Cambridge University Press, Cambridge, in preparation.

\bibitem{Litjens2018}
B.~Litjens.
\newblock Semidefinite bounds for mixed binary/ternary codes.
\newblock {\em Discrete Math.}, 341:1740--1748, 2018.

\bibitem{LitjensPS2017}
B.~Litjens, S.~Polak, and A.~Schrijver.
\newblock Semidefinite bounds for nonbinary codes based on quadruples.
\newblock {\em Des. Codes Cryptogr.}, 84:87--100, 2017.

\bibitem{Lovasz1979}
L.~Lov\'{a}sz.
\newblock On the {S}hannon capacity of a graph.
\newblock {\em IEEE Trans. Inform. Theory}, IT--25:1--7, 1979.

\bibitem{MartinT2009}
W.~J. Martin and H.~Tanaka.
\newblock Commutative association schemes.
\newblock {\em European J. Combin.}, 30:1497--1525, 2009.

\bibitem{McElieceRR1978}
R.~J. McEliece, E.~R. Rodemich, and H.~C. Rumsey, Jr.
\newblock The {L}ov\'{a}sz bound and some generalizations.
\newblock {\em J. Combin. Inform. System Sci.}, 3:134--152, 1978.

\bibitem{McElieceRRW1977}
R.~J. McEliece, E.~R. Rodemich, H.~Rumsey~Jr., and L.~Welch.
\newblock {New upper bounds on the rate of a code via the Delsarte--MacWilliams
  inequalities}.
\newblock {\em IEEE Trans. Inform. Theory}, IT--23:157--166, 1977.

\bibitem{MoralesT2018}
J.~V.~S. Morales and H.~Tanaka.
\newblock An {A}ssmus-{M}attson theorem for codes over commutative association
  schemes.
\newblock {\em Des. Codes Cryptogr.}, 86:1039--1062, 2018.

\bibitem{Nemirovski2007}
A.~Nemirovski.
\newblock Advances in convex optimization: conic programming.
\newblock In {\em International {C}ongress of {M}athematicians. {V}ol. {I}},
  pages 413--444. Eur. Math. Soc., Z\"{u}rich, 2007.

\bibitem{Polak2018}
S.~Polak.
\newblock Semidefinite programming bounds for {L}ee codes.
\newblock arXiv:1810.05066 [math.CO], 2018.

\bibitem{Polak2019}
S.~C. Polak.
\newblock {Semidefinite programming bounds for constant weight codes}.
\newblock {\em IEEE Trans. Inform. Theory}, IT--65:28--39, 2019.

\bibitem{Putinar1993}
M.~Putinar.
\newblock Positive polynomials on compact semi-algebraic sets.
\newblock {\em Indiana Univ. Math. J.}, 42:969--984, 1993.

\bibitem{Regts2009}
G.~Regts.
\newblock Upper bounds for ternary constant weight codes from semidefinite
  programming and representation theory.
\newblock Master's thesis, University of Amsterdam, 2009.

\bibitem{Schrijver1979}
A.~Schrijver.
\newblock {A comparison of the Delsarte and Lov\'asz bounds}.
\newblock {\em IEEE Trans. Inform. Theory}, IT--25:425--429, 1979.

\bibitem{Schrijver1986}
A.~Schrijver.
\newblock {\em Theory of linear and integer programming}.
\newblock John Wiley \& Sons, Ltd., Chichester, 1986.

\bibitem{Schrijver2005}
A.~Schrijver.
\newblock {New code upper bounds from the Terwilliger algebra and semidefinite
  programming}.
\newblock {\em IEEE Trans. Inform. Theory}, IT--51:2859--2866, 2005.

\bibitem{Shannon1956}
C.~E. Shannon.
\newblock The zero error capacity of a noisy channel.
\newblock {\em Institute of Radio Engineers, Transactions on Information
  Theory,}, IT-2:8--19, 1956.

\bibitem{Simon2011}
B.~Simon.
\newblock {\em {A course in convexity---An analytic viewpoint}}.
\newblock Cambridge University Press, Cambridge, 2011.

\bibitem{Srinivasan2011}
M.~K. Srinivasan.
\newblock Symmetric chains, {G}elfand-{T}setlin chains, and the {T}erwilliger
  algebra of the binary {H}amming scheme.
\newblock {\em J. Algebraic Combin.}, 34:301--322, 2011.

\bibitem{SudaT2014}
S.~Suda and H.~Tanaka.
\newblock A cross-intersection theorem for vector spaces based on semidefinite
  programming.
\newblock {\em Bull. Lond. Math. Soc.}, 46:342--348, 2014.

\bibitem{SudaTT2017}
S.~Suda, H.~Tanaka, and N.~Tokushige.
\newblock A semidefinite programming approach to a cross-intersection problem
  with measures.
\newblock {\em Math. Program.}, 166:113--130, 2017.

\bibitem{Tanaka2009}
H.~Tanaka.
\newblock New proofs of the {A}ssmus-{M}attson theorem based on the
  {T}erwilliger algebra.
\newblock {\em European J. Combin.}, 30:736--746, 2009.

\bibitem{Trinker2011}
H.~Trinker.
\newblock The triple distribution of codes and ordered codes.
\newblock {\em Discrete Math.}, 311:2283--2294, 2011.

\bibitem{Tuncel2010}
L.~Tun\c{c}el.
\newblock {\em Polyhedral and semidefinite programming methods in combinatorial
  optimization}.
\newblock American Mathematical Society, Providence, RI; Fields Institute for
  Research in Mathematical Sciences, Toronto, ON, 2010.

\bibitem{Vallentin2008}
F.~Vallentin.
\newblock {Lecture notes: Semidedfinite programming and harmonic analysis}.
\newblock arXiv:0809.2017 [math.OC], 2008.

\bibitem{Vallentin2009}
F.~Vallentin.
\newblock {Symmetry in semidefinite programs}.
\newblock {\em Linear Algebra and its Applications}, 430:360--369, 2009.

\bibitem{WolkowiczSV2000}
H.~Wolkowicz, R.~Saigal, and L.~Vandenberghe, editors.
\newblock {\em Handbook of semidefinite programming}.
\newblock Kluwer Academic Publishers, Boston, MA, 2000.

\bibitem{Wright1997}
S.~J. Wright.
\newblock {\em Primal-dual interior-point methods}.
\newblock Society for Industrial and Applied Mathematics (SIAM), Philadelphia,
  PA, 1997.

\end{thebibliography}
\end{document}